\documentclass[slac_one]{revtex4}
\usepackage{graphicx}
\usepackage{fancyhdr}
\usepackage{thumbpdf}
\usepackage[colorlinks=true, linkcolor=blue, citecolor=blue, urlcolor=blue,
pdfauthor={Philip Bechtle, Klaus Desch, Peter Wienemann},
pdftitle={Supersymmetry Parameter Analysis with Fittino},
pdfkeywords={SUSY SPA Fittino}]{hyperref}
\usepackage{amsmath,amssymb,dcolumn,afterpage,supertabular}
\newcolumntype{e}[1]{D{.}{.}{#1}}
\begin{document}

%Title of paper
\title{{\small{2005 International Linear Collider Workshop - Stanford,
U.S.A.}\hfill{\normalfont SLAC-PUB-11295}}\\ %% Please keep this conference title here
\vspace{12pt}
Supersymmetry Parameter Analysis with Fittino} %% Paper title goes here

% Repeat the \author .. \affiliation  etc. as needed
%
% \affiliation command applies to all authors since the last
% \affiliation command. The \affiliation command should follow the
% other information

\author{P. Bechtle}
\affiliation{Stanford Linear Accelerator Center (SLAC),
  2575 Sand Hill Road, Menlo Park, CA 94025, USA}
\author{K. Desch}
\affiliation{Universit\"at Freiburg, Physikalisches Institut,
	 Hermann-Herder-Str.~3,
	 D-79104 Freiburg, Germany}
\author{P. Wienemann}
\affiliation{Universit\"at Freiburg, Physikalisches Institut,
	 Hermann-Herder-Str.~3,
	 D-79104 Freiburg, Germany}

\begin{abstract}
We present the results of a realistic global fit of the Lagrangian
parameters of the Minimal Supersymmetric Standard Model to simulated
data from ILC and LHC with realistic estimates of the observable
uncertainties.  Higher order radiative corrections are accounted for
where ever possible to date.  Results are obtained for a modified SPS1a
MSSM benchmark scenario but they were checked not to depend critically
on this assumption. Exploiting a simulated annealing algorithm,
a stable result is obtained without any {\it a priori} assumptions on the
fit parameters. Most of the Lagrangian parameters can be extracted at
the percent level or better if theoretical uncertainties are
neglected. Neither LHC nor ILC measurements alone will be sufficient
to obtain a stable result. The effects of theoretical uncertainties
arising from unknown higher-order corrections and parametric 
uncertainties are examined qualitatively. They appear to be relevant
and the result motivates further precision calculations.
\end{abstract}

%\maketitle must follow title, authors, abstract
\maketitle

\thispagestyle{fancy}

%====================================================================
\section{INTRODUCTION} \label{sec:Intro}
If low-energy Supersymmetry (SUSY)~\cite{susy} is realized in nature,
the next generation of colliders, the Large Hadron Collider
(LHC) and the International Linear Collider
(ILC) are likely to copiously produce SUSY particles and
will allow for precise measurements of their properties. Once SUSY is
established experimentally, it is the main task to explore the unknown
mechanism of SUSY breaking (SSB). The most general way to do so is to
reconstruct the parameters of the general Minimal Supersymmetric
Standard Model (MSSM)~\cite{mssm} parameter space, which is
significantly more ambitious than a test of specific models of SUSY
breaking, like e.~g.~minimal supergravity (mSUGRA)~\cite{msugra} can
be tested against the data in a relative straight-forward manner (see
e.~g.~\cite{msugrafits}) due to their small number of parameters.

The general MSSM Lagrangian contains more than 100 new parameters
which are only related to each other through the unknown SSB
mechanism. Many of them (like complex phases) are already limited to
some extent by measurements, e.g. by the absence of neutron and
electron electric dipole moments~\cite{edm}. Furthermore universality
of first and second generation appears to be a reasonable
approximation while large Yukawa couplings in the third generation lead to
significant differences. Applying these constraints, the number of
free parameters is reduced to 19, including the Standard Model top
quark mass as a parameter to account for parametric uncertainties. It
is of high interest how well this 19-dimensional parameter space can
be restricted with future measurements from LHC and ILC.

The experimental collaborations ATLAS~\cite{atlas} and CMS~\cite{CMS}
at the LHC have performed detailed simulations of the possibilities to
extract mass information from their future data. At a future electron
positron collider for collision energies up to 1~TeV, the
International Linear Collider (ILC), the kinematically accessible part
of the superpartner spectrum can be studied in great detail due to
favorable background conditions and the well-known initial
state~\cite{ref:LHCILC}. The attempt of an evaluation of the MSSM
Lagrangian parameters and their associated errors is only useful if
the experimental errors of the future measurements are known and under
control. The by far best-studied SUSY scenario to serve as a basis for
such an evaluation is an mSUGRA-inspired benchmark scenario, the SPS1a
scenario~\cite{ref:SPS}.  For this scenario with a relatively light
superpartner spectrum, a wealth of experimental simulations exists and
has recently been compiled in the framework of the international
LHC/ILC study group~\cite{ref:LHCILC}. In this paper, we perform a
global fit of the 19 MSSM parameters to those expected measurements
augmented by possible measurements of topological cross-sections
(i.~e.~cross sections times branching fractions) that will be possible
at the ILC with polarized beams. Since a detailed experimental
simulation for some measurements is lacking, we estimate their
uncertainties conservatively from the predicted cross-sections and
transferring experimental efficiency from well-studied cases.

As theoretical basis for the global fit presented in this paper we use
the loop-level calculations as implemented into the program
SPheno~\cite{ref:SPheno}.  In SPheno, the masses and decay branching
fractions of the superpartners are calculated as well as production cross
sections in $\text{e}^+\text{e}^-$ collisions.
 
In this paper we present the result of a global fit of the MSSM
Lagrangian parameters at the electro-weak scale for a slightly
modified SPS1a benchmark scenario using the program
Fittino~\cite{ref:Fittino,ref:FittinoProgram}. Previous evaluations of
the errors of those parameters~\cite{msugrafits} did not attempt to
develop a strategy to extract those parameters from data without {\it
a priori} knowledge. Within Fittino, special attention is given
precisely to this task, i.~e.~to find the parameter set which is most
consistent with the data in a $\chi^2$ minimization before a careful
evaluation of errors and correlations is performed. A similar
code~\cite{sfitter} exploits a different strategy for this task.

This paper is organized as follows. In Section~\ref{sec:Fit} we
shortly describe the approach used in Fittino. In
Subsection~\ref{sec:Inputs}, the input observables are explained. The
results of the fit and the error evaluation method are summarized in
Subsection~\ref{sec:Results} and conclusions are drawn in
Section~\ref{sec:conclusions}.

%====================================================================
\section{SPS1a' FIT} \label{sec:Fit}
From the numerous fitting options provided by Fittino we have chosen
the following fit procedure to extract the low-energy Lagrangian
parameters. First, in order to be independent of human bias, start
values for the parameters are calculated using tree-level relations
between parameters and a few
observables~\cite{ref:FittinoProgram}. For fits with many parameters
these values are not good enough to allow a fitting tool like
MINUIT~\cite{ref:Minuit} to find the global minimum due to the amount
of loop-level induced cross-dependencies between the individual
sectors of the MSSM. Therefore, in a second step, the parameters are
refined, using a simulated annealing
approach~\cite{ref:FittinoProgram,SimAnn,SimAnn2}. As a result, the
parameter values are close to the global minimum so that a global fit
using MINUIT can find the exact minimum in a third step. To determine
the parameter uncertainties and correlations many individual fits with input
values randomly smeared within their uncertainty range are carried
out. The parameter uncertainties and the correlation
matrix are derived from the spread of the fitted parameter
values.

%====================================================================
\subsection{Input Observables from LHC and ILC} \label{sec:Inputs}
A number of anticipated LHC and ILC measurements serve as input
observables to the fit.  For the ILC, running at center-of-mass
energies of 400 GeV, 500 GeV and 1 TeV is considered with 80~\%
electron and 60~\% positron polarization.  The predicted values of the
observables are calculated using the following prescriptions:
\begin{itemize}
\item {\bf Masses:} \newline The experimental uncertainties of the
  mass measurements are taken from \cite{ref:LHCILC}. 
\item {\bf Cross-sections:} \newline Only $\text{e}^+ \text{e}^-$
  cross-sections are included in the fit. However, the measurement of
  absolute cross-sections is impossible for many channels, in which
  only a fraction of the final states can be reconstructed. Therefore,
  absolute cross-section measurements are only used for the
  Higgs-strahlung production of the light Higgs boson, which is
  studied in detail in~\cite{ref:HiggsBr}.
\item {\bf Cross-sections times branching fractions:} \newline Since
  no comprehensive study of the precision of cross-section times
  branching fraction measurements is available, the uncertainty is
  assumed to be the error of a counting experiment with the following
  assumptions:
  \begin{itemize}
  \item The selection efficiency amounts to 50~\%.
  \item 80~\% polarization of the electron beam and 60~\%
    polarization of the positron can be achieved.
  \item 500~fb$^{-1}$ per center-of-mass energy and polarization
    is collected.
  \item The relative precision is not allowed to be better than 1~\%
    and the absolute accuracy is at most 0.1~fb to account for
    systematic uncertainties.
  \end{itemize}
  All production processes and decays of SUSY particles and Higgs
  bosons are used which have a cross-section times branching ratio
  value of more than 1~fb in one of the $\text{e}^+ \text{e}^-$
  polarization states LL, RR, LR, RL at $\sqrt{s}=400$ and $500$~GeV
  and LR or RL at $\sqrt{s}=1000$~GeV.
\item {\bf Branching fractions:} \newline The three largest branching
  fractions of the lightest Higgs boson are included. The
  uncertainties on the Higgs branching fractions are taken
  from~\cite{ref:HiggsBr}.
\item {\bf Standard Model parameters:} \newline The present
  uncertainties for $m_{\text{W}}$ and $m_{\text{Z}}$ are used as
  conservative estimates. The experimental uncertainty for the top
  mass $m_{\text{t}}$ is assumed to be 50 MeV~\cite{ref:HiggsBr}.
\end{itemize}

\begin{table}[t]
\caption{The Fittino fit result for the SPS1a' motivated scenario.
  The left column shows the values predicted by SPheno version 2.2.2
  for this scenario, the second column exhibits the central values of
  the fit with unsmeared input observables, the third column displays
  the parameter uncertainties for the fit with experimental
  uncertainties only.  The last column shows the parameter
  uncertainties for the fit with experimental and theoretical
  uncertainties.\vspace{2mm}}
\label{tab:UnsmearedFitResultSPS1aPrime}
\begin{center}
\begin{tabular}{l e{7} e{7} e{7} e{7}}
\hline
\multicolumn{1}{c}{Parameter} & \multicolumn{1}{c}{``True'' value} & 
\multicolumn{1}{c}{Fit value} & \multicolumn{1}{r}{Uncertainty} &
\multicolumn{1}{r}{Uncertainty} \\
 & & & \multicolumn{1}{c}{(exp.)} & \multicolumn{1}{c}{(exp.+theor.)}\\
 \hline
$\tan\beta$          &  10.00                &  10.00                 & 0.11 & 0.15 \\
$\mu$                &  400.4  \;\text{GeV}  &  400.4   \;\text{GeV}  & 1.2\;\text{GeV} & 1.3\;\text{GeV} \\
$X_{\tau}$           &  -4449.  \;\text{GeV}  & -4449.    \;\text{GeV}  & 20.\;\text{GeV} & 30. \;\text{GeV} \\
$M_{\tilde{e}_R}$    &  115.60  \;\text{GeV} &  115.60 \;\text{GeV}   &  0.27 \;\text{GeV} & 0.50   \;\text{GeV} \\
$M_{\tilde{\tau}_R}$ &  109.89  \;\text{GeV} &  109.89   \;\text{GeV} &  0.41  \;\text{GeV} & 0.60    \;\text{GeV} \\
$M_{\tilde{e}_L}$    &  181.30  \;\text{GeV} &  181.30   \;\text{GeV} &  0.10  \;\text{GeV} & 0.12    \;\text{GeV} \\
$M_{\tilde{\tau}_L}$ &  179.54  \;\text{GeV} &  179.54   \;\text{GeV} &  0.14  \;\text{GeV} & 0.19    \;\text{GeV} \\
$X_{\text{t}}$     &  -565.7  \;\text{GeV} & -565.7    \;\text{GeV} & 3.1   \;\text{GeV} & 15.4   \;\text{GeV} \\
$X_{\text{b}}$  &  -4935. \;\text{GeV} & -4935.   \;\text{GeV} & 1284. \;\text{GeV} & 1825.  \;\text{GeV} \\
$M_{\tilde{q}_R}$    &  503.   \;\text{GeV} &   503.   \;\text{GeV} & 24.  \;\text{GeV} &  27.  \;\text{GeV} \\
$M_{\tilde{b}_R}$    &  497.   \;\text{GeV} &  497.    \;\text{GeV} & 8.  \;\text{GeV} &  15.  \;\text{GeV} \\
$M_{\tilde{t}_R}$    &  380.9   \;\text{GeV} &  380.9    \;\text{GeV} & 2.5  \;\text{GeV} &  3.9    \;\text{GeV} \\
$M_{\tilde{q}_L}$    &  523.   \;\text{GeV} &  523.    \;\text{GeV} &  10.  \;\text{GeV} & 15.    \;\text{GeV} \\
$M_{\tilde{t}_L}$    &  467.7   \;\text{GeV} &  467.7    \;\text{GeV} & 3.1 \;\text{GeV} & 5.1    \;\text{GeV} \\
$M_1$                &  103.27 \;\text{GeV} &  103.27  \;\text{GeV} & 0.06  \;\text{GeV} & 0.14   \;\text{GeV} \\
$M_2$                &  193.45 \;\text{GeV} &  193.45  \;\text{GeV} & 0.10  \;\text{GeV} & 0.15  \;\text{GeV} \\
$M_3$                &  569.  \;\text{GeV} &  569.    \;\text{GeV} &  7.  \;\text{GeV} &   7.   \;\text{GeV} \\
$m_{\text{A}_{\text{run}}}$ & 312.0 \;\text{GeV} & 311.9 \;\text{GeV} & 4.6  \;\text{GeV} & 6.9   \;\text{GeV} \\
$m_{\text{t}}$     &  178.00  \;\text{GeV} &  178.00   \;\text{GeV} &  0.050  \;\text{GeV} & 0.108    \;\text{GeV} \\
\hline
\multicolumn{5}{c}{$\chi^2$ for unsmeared observables: $5.3\times 10^{-5}$}\\
\hline
\end{tabular}
\end{center}
\end{table}

In order to check the influence of theoretical uncertainties on the
fit results, the fit has been performed twice, once with experimental
uncertainties only and a second time with experimental and theoretical
uncertainties. Theoretical uncertainties for the mass predictions are
scale uncertainties taken from~\cite{ref:SPA}. The experimental and
theoretical contributions are added in quadrature. The assumed
experimental and theoretical uncertainties for the masses are given in
\cite{ref:LHCILC}. For the cross-section and cross-section times
branching fraction measurements, the smallest allowed relative
precision is raised to 2 \% for the fit including theoretical
uncertainties (as opposed to 1 \% for the fit without theoretical
uncertainties). The full list of observables used in the fit and their
uncertainties can be obtained from~\cite{ref:Fittino}.

\subsection{Fit Results} \label{sec:Results}
The input observables described in Section~\ref{sec:Inputs} are used
to determine the SUSY Lagrangian parameters in a global fit under the
assumptions mentioned in Section~\ref{sec:Intro}. In total 18 SUSY
parameters remain to be fitted. In addition to those, the top mass
$m_{\text{t}}$ is fitted, since it strongly influences parts of the
MSSM observables.  Thus 19 Lagrangian parameters are simultaneously
determined in this fit.

As shown in Table~\ref{tab:UnsmearedFitResultSPS1aPrime} all
parameters are perfectly reconstructed at their input values. Due to
the fact that the input observables are unsmeared in this fit, the
final $\chi^2$ is close to zero at $\chi^2 = 5.3 \times 10^{-5}$.

Most parameters are reconstructed to a precision better than or around
1~\%. For the U(1) gaugino mass parameter $M_1$, an accurary below the
per mil level is achieved for the fit including experimental
uncertainties only.  However, its precision is worse by a factor of 2
once theoretical uncertainties are included. The uncertainties on
other parameters increase by up to a factor of 5 if theoretical
uncertainties are taken into account, such as $X_{\text{t}}$. In order
to fully benefit from the precision data provided in particular by the
ILC, work must be invested to reduce uncertainties of theoretical
predications.

\begin{figure*}[t]
\centering{
\includegraphics[width=0.45\textwidth]{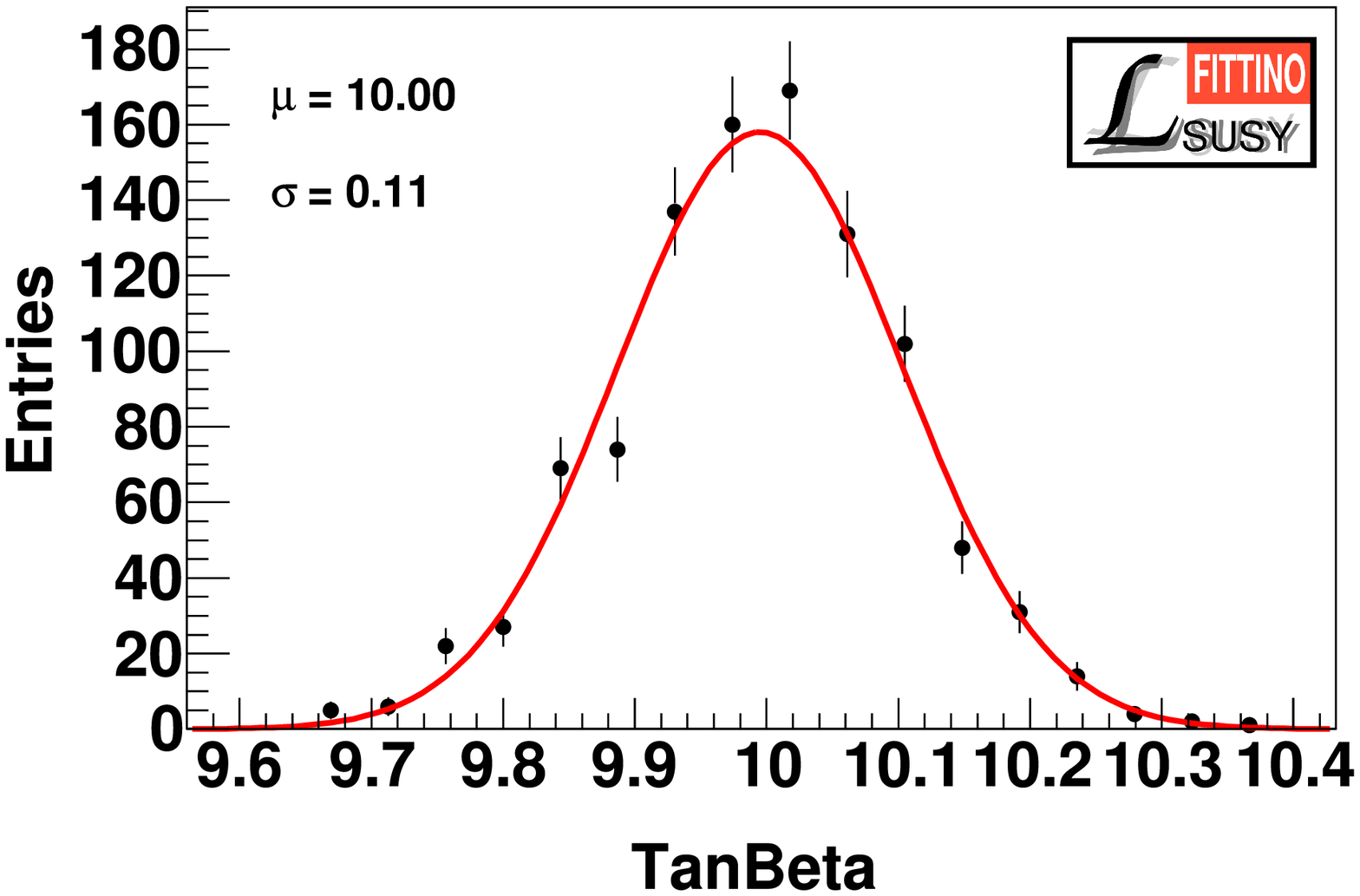}
\includegraphics[width=0.45\textwidth]{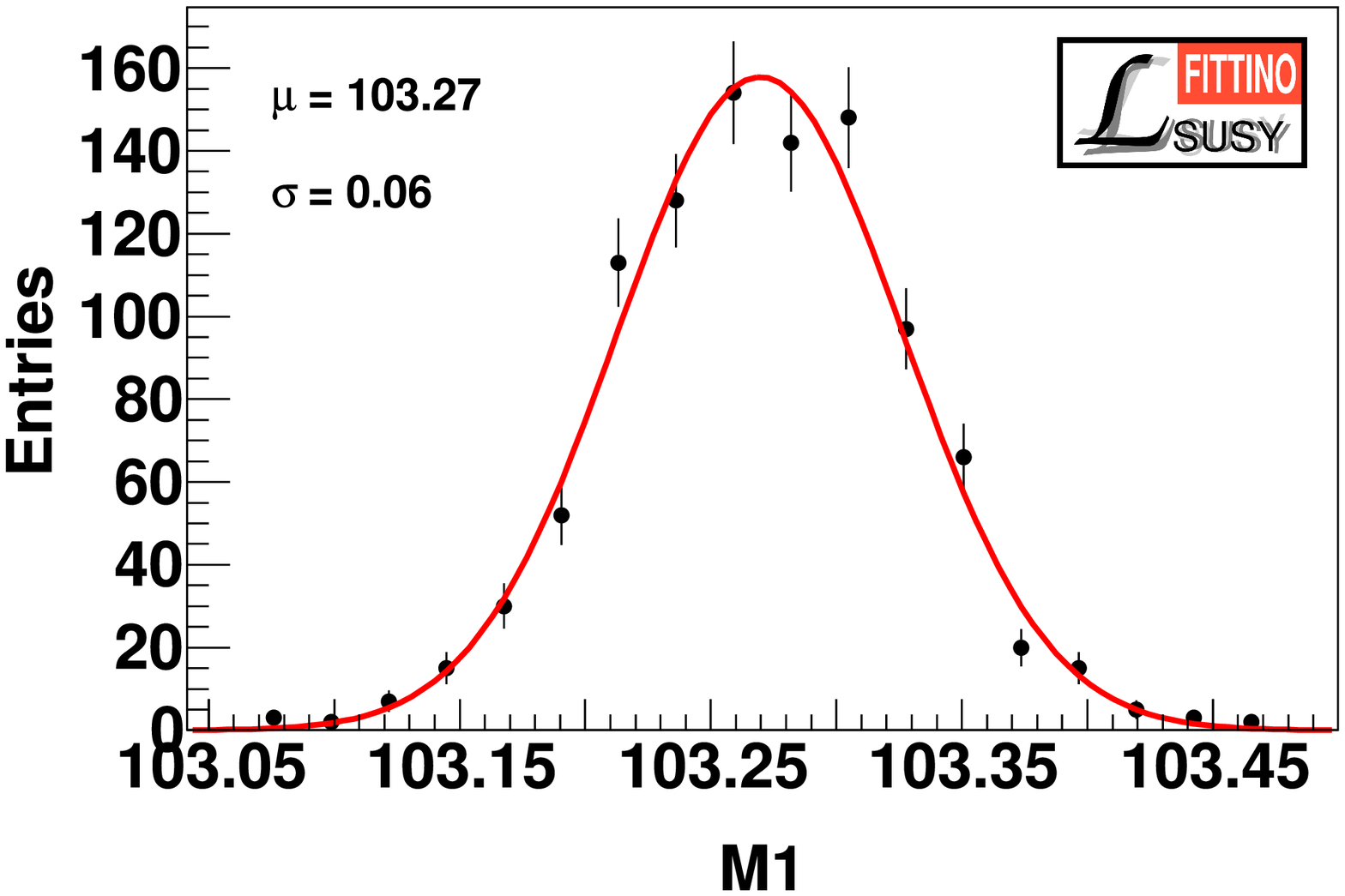}
}
\caption{Examples for the parameter distributions from the toy
fits. No theoretical uncertainties are included. The agreement of the
parameter distributions with the Gaussian hypothesis is very good.}
\label{pulldistributions}
\end{figure*}

After a successful convergence of the simulated annealing algorithm to
the input parameter values, the fit uncertainties are evaluated by
carrying out many individual fits with input values randomly smeared
within their uncertainty range using a Gaussian probability
density. The complete covariance matrix and the correlation matrix are
derived from the spread of the fitted parameter values.  For the fit
without theoretical uncertainties 1002 such fits were made. The
corresponding number for the case including theory uncertainties
amounts to 993. For each of those fits, the parameter set is
minimized. For a large and complex parameter space with large
correlations among the parameters this method has turned out to be
more robust than a MINOS error analysis.  An example of the outcome of
this procedure for the fit (with experimental uncertainties only) is
shown in Figure~\ref{pulldistributions} for the parameters $\tan\beta$
and $M_1$ for the 1002 independent fits.

%\begin{figure}[t]
%    \begin{center}
%      \includegraphics[width=0.9\textwidth]{figures/relunc_ILCLHC}
%    \end{center}
%\caption{Relative uncertainties 
%	of the parameter measurements with (red) and without (blue) 
%	theoretical uncertainties.}
%\label{fig:RelativeUncertainties}
%\end{figure}

\begin{figure*}[t]
\centering
\includegraphics[width=0.5\textwidth]{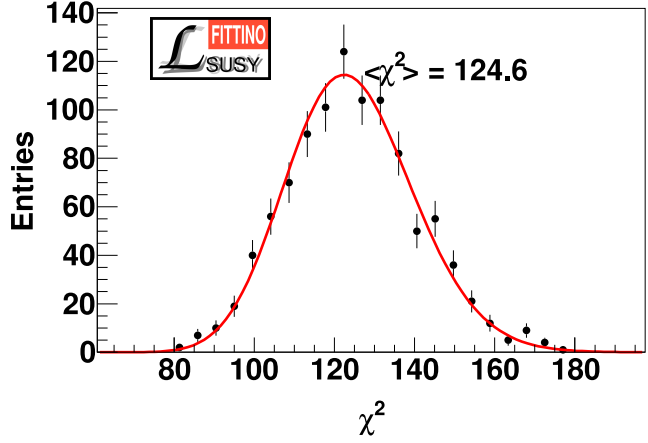}
\caption{The $\chi^2$ distribution obtained from 1002 toy fits with
smeared input distributions. The agreement with the expectation of a
$\chi^2$ distribution with a mean of $126\pm 0.64$ is very good.}
\label{chisqdistribution}
\end{figure*}

Figure~\ref{chisqdistribution} shows the distribution of the $\chi^2$ values
for these 1002 independent fits for the case without theoretical
uncertainties.  The mean $\chi^2$ obtained from a fit of the $\chi^2$
distribution to the observed distribution is 124.6, in agreement with
the expectation of 126~$\pm$~0.64. This shows that the fits converge
well to the true minimum of the $\chi^2$ surface for the smeared
observables, implying that the uncertainties extracted from the toy
fit value distributions are correct.

%====================================================================
\subsection{Relative Importance of Observables}
In order to determine automatically the most important input
observables influencing the precision with which a certain parameter,
the contribution of each observable to $\Delta \chi^2 =
\chi^2_{1\sigma} - \chi^2_{\text{min}}$ is calculated for the
parameter of interest.  $\chi^2_{\text{min}}$ is the $\chi^2$ value at
the minimum of the fit and $\chi^2_{1\sigma}$ is the corresponding
value if the parameter is varied within $\pm 1 \sigma$.  A large
$\Delta \chi^2$ indicates a strong correlation of the parameter with
other parameters. A large contribution $\Delta \chi^2_i$ of an
observable $i$ to the total $\Delta \chi^2$ with respect to the
contributions by the other observables means that the parameter is
mainly determined by the measurement of observable $i$.

\begin{table}[t]
\caption{Individual $\Delta\chi^2$ contributions in the Fittino
    SPS1a' motivated fit. In the first column the parameter and its fitted value
    is shown.  The second column shows the total $\Delta\chi^2$ for
    this parameter, if the parameter is varied by $1\,\sigma$. The
    third and fourth column show the parameters and their contribution
    to the $\Delta\chi^2$ of the parameter in per cent.\vspace{2mm}}\label{tab:IndividualChi2}
\begin{center}
\begin{tabular}{l e{5} l e{7}}
\hline
    Parameter & \multicolumn{1}{c}{Total $\Delta\chi^2$} & Observable & \multicolumn{1}{c}{Contribution to the}  \\
    Value     &                                          &            & \multicolumn{1}{c}{$\Delta\chi^2$ in \%} \\
\hline
\hline
    $\tan\beta$              &  5.0 & $\sigma(\text{e}^-_L\text{e}^+_R
                                       \rightarrow\mathrm{H}^{\pm}\mathrm{H}^{\mp}
                                       \rightarrow\mathrm{t}\bar{\mathrm{b}}
                                       \bar{\mathrm{t}}\mathrm{b})~1$~TeV                 & 31.1 \\
    $10.00 \pm 0.11$         &      & $\sigma(\text{e}^-_L\text{e}^+_R
                                       \rightarrow\mathrm{H}\mathrm{A}
                                       \rightarrow\mathrm{b}\bar{\mathrm{b}}
                                       \mathrm{b}\bar{\mathrm{b}}~1)$~TeV               & 9.61 \\
                             &      & $m_{\mathrm{h}}$         & 8.12 \\
    \hline                      
    %%%%%%%%%%%%%%%%%%%%%%%%%%%%%%%%%%%%%%%%%%%%%%%%%%%%%%%%%%%%%%%%%%%%%%%%%%%%%%%%%%%%%%%%%%%%%%
    $\mu$                    & 15.2 & $\sigma(\mathrm{e}^-_L\mathrm{e}^+_R
                                       \rightarrow\tilde{\chi}^+_1\tilde{\chi}^-_1
                                       \rightarrow\bar{\nu}_{\tau}\tilde{\chi}^0_1{\tau}^{+}\,
                                       \nu_{\tau}\tilde{\chi}^0_1{\tau}^{-})~400$~GeV   & 14.5 \\
    $400.39 \pm 1.18$~GeV    &      & $\sigma(\mathrm{e}^-_L\mathrm{e}^+_R
                                       \rightarrow\tilde{\chi}^+_1\tilde{\chi}^-_1
                                       \rightarrow\bar{\nu}_{\tau}\tilde{\chi}^0_1{\tau}^{+}\,
                                       \nu_{\tau}\tilde{\chi}^0_1{\tau}^{-})~500$~GeV             & 7.49 \\
                             &      & $\sigma(\mathrm{e}^-_R\mathrm{e}^+_R
                                       \rightarrow\tilde{\chi}^+_1\tilde{\chi}^-_1
                                       \rightarrow\bar{\nu}_{\tau}\tilde{\chi}^0_1{\tau}^{-}\,
                                       \nu_{\tau}\tilde{\chi}^0_1{\tau}^{+})~500$~GeV    &  6.71 \\
    \hline
    $M_{\tilde{\mathrm{q}}_R}$ & 36.6 & $m_{\tilde{\mathrm{e}}_R}$ & 53.6 \\
    $501.6 \pm 23.6$~GeV     &      & $m_{\tilde{\mu}_R}$      &  3.34 \\
                             &      & $\sigma(\mathrm{e}^-_L\mathrm{e}^+_R
                                       \rightarrow\tilde{\chi}^+_1\tilde{\chi}^-_1
                                       \rightarrow\bar{\nu}_{\tau}\tilde{\chi}^0_1{\tau}^{-}\,
                                       \nu_{\tau}\tilde{\chi}^0_1{\tau}^{+})~1$~TeV &  3.25 \\
    \hline
\end{tabular}
\end{center}
\end{table}

Generally the precision of most parameters is determined by ILC data,
as shown as an example for $\tan\beta$ and $\mu$ in
Table~\ref{tab:IndividualChi2}. As a surprise it can be seen that even
for the uncertainty of the squark mass parameters the most
constraining input is provided by ILC slepton mass measurements. It
shows that the small contribution of the squark mass parameter
$M_{\tilde{\text{q}}_{R}}$ to the slepton mass $m_{\tilde{e}_R}$ has a
larger effect compared to the good precision of the measurement of
$m_{\tilde{e}_R}$, than the large contribution of the squark mass
parameter to the much less precisely measured squark mass.

%====================================================================
\subsection{LHC and ILC Synergy}
In order to test the dependence of the fit results on the synergy of
LHC and ILC, a fit has been performed with all ILC-influenced
observables removed. While the LHC observables may allow to determine
a further reduced set of parameters (e.g. mSUGRA, if a simplified
model described by fewer parameters still fits the data), the complete
set of parameters studied here is not constrained by the LHC mass
observables which currently enter the fit. For most parameters, the
resulting parameter uncertainties exceed the LHC+ILC uncertainties by
orders of magnitude. Only for 3 of the parameters studied here, namely
$M_{\tilde{\mathrm{q}}_L}$, $M_{\tilde{\mathrm{q}}_R}$ and $M_3$, the
uncertainty is in the same order of magnitude. This exemplifies the
synergy effects of the LHC and the ILC.

%\begin{figure*}[t]
%\centering
%\includegraphics[width=0.8\textwidth]{figures/propaganda}
%\caption{Relative size of the relative uncertainties of the parameter measurements 
%  at LHC only and at LHC+ILC.} \label{propaganda}
%\end{figure*}

%====================================================================
\section{CONCLUSIONS} \label{sec:conclusions}
The Lagrangian parameters of the MSSM assuming universality for the
first and second generation and real parameters but without
assumptions on the SUSY breaking mechanism are correctly reconstructed
without usage of {\it a priori} information. This has been achieved using
precision measurements at the LHC and ILC as input to a global fit
exploiting the techniques implemented in the program Fittino.

Most of the Lagrangian parameters can be determined to a precision
around the percent level. For some parameters an accuracy of better
than 1~per mil is achievable if theoretical uncertainties are
neglected. A fit including them revealed that theory uncertainties can
significantly deteriorate the precision of the Lagrangian parameter
determination. More work is needed to improve the accuracy of
theoretical predictions in order to fully benefit from the
experimental precision. The only parameter which cannot be strongly
constrained by the observables used in the presented fit is
$X_{\text{b}}$ resulting in a precision of only 30 \% for this
parameter.

%====================================================================
% If you have acknowledgments, this puts in the proper section head.
\begin{acknowledgments}
The authors are grateful to Gudrid Moortgat-Pick, Werner Porod, Sven
Heinemeyer and the whole SPA working group for very fruitful
discussions.
\end{acknowledgments}

%====================================================================

\end{document}